# Observation of perfect diamagnetism and interfacial effect on the electronic structures in infinite layer $Nd_{0.8}Sr_{0.2}NiO_2$ superconductors

S. W. Zeng,[1, #,*] X. M. Yin,[2,3,#] C. J. Li,[4,5,#], L. E. Chow,[1,#] C. S. Tang,[2,6] K. Han,[1,7] Z. Huang,[1,7] Y. Cao,[8] D. Y. Wan,[1] Z. T. Zhang,[1] Z. S. Lim,[1] C. Z. Diao,[2] P. Yang,[2,4] A. T. S. Wee,[1,2] S. J. Pennycook,[4] A. Ariando,[1,*]

[1]*Department of Physics, Faculty of Science, National University of Singapore, Singapore 117551, Singapore*

[2]*Singapore Synchrotron Light Source (SSLS), National University of Singapore, 5 Research Link, Singapore 117603, Singapore*

[3]*Shanghai Key Laboratory of High Temperature Superconductors, Physics Department, Shanghai University, Shanghai 200444, China*

[4]*Department of Materials Science and Engineering, National University of Singapore, Singapore 117575, Singapore*

[5]*Department of Materials Science and Engineering, Southern University of Science and Technology, Shenzhen 518055, Guangdong, China*

[6]*Institute of Materials Research and Engineering, A\*STAR (Agency for Science, Technology and Research), 2 Fusionopolis Way, Singapore, 138634 Singapore*

[7]*Information Materials and Intelligent Sensing Laboratory of Anhui Province, Institutes of Physical Science and Information Technology, Anhui University, Hefei 230601, Anhui, China*

[8]*Department of Electrical and Computer Engineering, National University of Singapore, Singapore 117583, Singapore*

#The authors contributed equally to this work.

*To whom correspondence should be addressed: ariando@nus.edu.sg

**Nickel-based complex oxides have served as a playground for decades in the quest for a copper-oxide analog of the high-temperature superconductivity. They may provide clues towards understanding the mechanism and an alternative route for high-temperature superconductors. The recent discovery of superconductivity in the infinite-layer nickelate thin films has fulfilled this pursuit. However, material synthesis remains challenging, direct demonstration of perfect diamagnetism is still missing, and understanding of the role of the interface and bulk to the superconducting properties is still lacking. Here, we show high-quality $Nd_{0.8}Sr_{0.2}NiO_2$ thin films with different thicknesses and demonstrate the interface and strain effects on the electrical, magnetic and optical properties. Perfect diamagnetism is achieved, confirming the occurrence of superconductivity in the films. Unlike the thick films in which the normal-state Hall-coefficient changes signs as the temperature decreases, the Hall-coefficient of films thinner than 5.5 nm remains negative, suggesting a thickness-driven band structure modification. Moreover, X-ray absorption spectroscopy reveals the Ni-O hybridization nature in doped infinite-layer nickelates, and the hybridization is enhanced as the thickness decreases. Consistent with band structure calculations on the nickelate/$SrTiO_3$ heterostructure, the interface and strain effect induce a dominating electron-like band in the ultrathin film, thus causing the sign-change of the Hall-coefficient.**

**INTRODUCTION**

The search for the nickelate superconductivity was enthused by the idea of mimicking the $3d_{x^2-y^2}$ orbital of the single-band high-$T_c$ cuprates[1]. However, the recent results indicate the complex multiband structures in doped infinite-layer nickelates, suggesting a new family of superconductivity[1,2]. Numerous theoretical works have been conducted based on the bulk pictures[2]; however, the superconductivity has only been observed in epitaxial $Nd_{1-x}Sr_xNiO_2$ and $Pr_{1-x}Sr_xNiO_2$ ultrathin films (up to ~10 nm) with an infinite-layer structure[1,3-9]. In contrast, infinite-layer nickelates prepared in bulk form show only insulating behavior[10-12]. Moreover, the DC diamagnetic response in the superconducting thin films has never been demonstrated[1]. These beg the question of whether the superconductivity occurs in the whole film or at the interface between the nickelate and $SrTiO_3$ (STO) substrate[13-16]. Theoretical calculation further proposed that the interface/surface-induced Fermi surface modification causes the transformation from a $d$-wave paring in bulk into an $s$-wave paring at the interface/surface[2,17], which might be consistent with the recent observation of two gaps from the tunneling spectrum measurement[5]. This further adds to the puzzle of whether the observed electronic properties are associated with the bulk or the heterostructure interface. In this work, we comprehensively investigate the $Nd_{0.8}Sr_{0.2}NiO_2$ films of various thicknesses to confirm the bulk nature of the superconductivity and reveal the interfacial effects on the multiband picture of the infinite-layer nickelate thin films and demonstrate their perfect DC diamagnetic response.

## RESULTS

**Infinite-layer Structure**

Figure 1a shows the X-ray diffraction (XRD) $\theta$–$2\theta$ patterns of the $Nd_{0.8}Sr_{0.2}NiO_2$ thin films of different thicknesses from 4.6 to 15.2 nm. The XRD characterization of the as-grown perovskite $Nd_{0.8}Sr_{0.2}NiO_3$ thin films can be found in Supplementary Fig. 1. The obvious diffraction peak and thickness oscillations (Laue fringes) in the vicinity of the ($00l$) infinite-layer peak ($l$ is an integer) indicate the high crystallinity of the films. The ($00l$) peak positions slightly shift towards a higher angle as the thickness increases, indicating a shrinking of the $c$-axis, with the lattice constants $c$ change from ~3.42 Å for the 4.6-nm film to ~3.36 Å for the 15.2-nm film, as plotted in Fig. 1b.

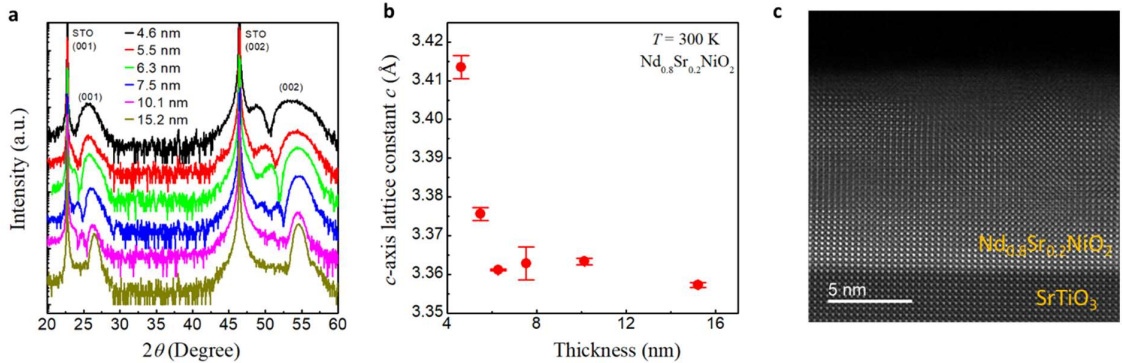

**Figure 1: Thickness dependence of the infinite layer structure.** (a) The XRD $\theta$–$2\theta$ scan patterns of the $Nd_{0.8}Sr_{0.2}NiO_2$ thin films with different thicknesses on $SrTiO_3$ substrates. The intensity is vertically displaced for clarity. (b) The room-temperature $c$-axis lattice constants, $c$, as a function of thickness, as calculated from the (001) peak positions. The red dots represent the average $c$ of the two sets of samples and the error bars represent the variance. (c) The HAADF-STEM image of the 10.1-nm $Nd_{0.8}Sr_{0.2}NiO_2$ on $SrTiO_3$ substrate.

It has been shown that $Nd_{1-x}Sr_xNiO_2$ ($x \leq 0.2$) bulks exhibit an in-plane lattice constant (3.914 − 3.921 Å) slightly larger than that (3.905 Å) of STO substrate[10,11], suggesting the presence of compressive strain on the films imposed by the substrate. The extent of compressive strain decreases as the film thickness increases and thus causing the shrinking of $c$. Figure 1c shows the high-angle annular dark-field scanning transmission electron microscopy (HAADF-STEM) image of the 10.1-nm film. A clear infinite-layer structure is observed with no obvious defect throughout the layer.

**Electronic Properties**

The resistivity versus temperature ($\rho$-$T$) curves for the $Nd_{0.8}Sr_{0.2}NiO_2$ thin films are shown in Fig. 2a, and the zoomed-in $\rho$-$T$ curves at temperatures from 50 to 2 K are shown in the inset. All the samples behave like a metal at the normal state and are superconducting at low temperatures. The onset superconducting transition temperature $T_{c,90\%R}$ (defined as the temperature at which the resistivity drops to 90% of the value at 15 K) and zero-resistance $T_{c,zero-R}$ decrease with decreasing thickness (Fig. 2f). Figure 2b shows the corresponding temperature dependence of the normal-state Hall coefficients ($R_H$) of the $Nd_{0.8}Sr_{0.2}NiO_2$ films. The $R_H$ for samples with a thickness higher than 6.8 nm shows a negative sign at room temperature and undergoes a smooth transition to a positive sign at a low temperature of ~50 K, consistent with previous observation at the doping level $x = 0.2$[1,3,4]. However, as the thickness decreases to 6.8 nm, the $R_H$ sign-change temperature decreases to 22 K. The $R_H$ even remains negative at the whole temperature range below 300 K for the films with thickness lower than 6.8 nm. Figure 2e presents the thickness dependence of the $R_H$ at 20 K and 300 K, clearly showing a sign-change from positive to negative with decreasing thickness. This suggests a change of the multiband structures upon reducing thickness.

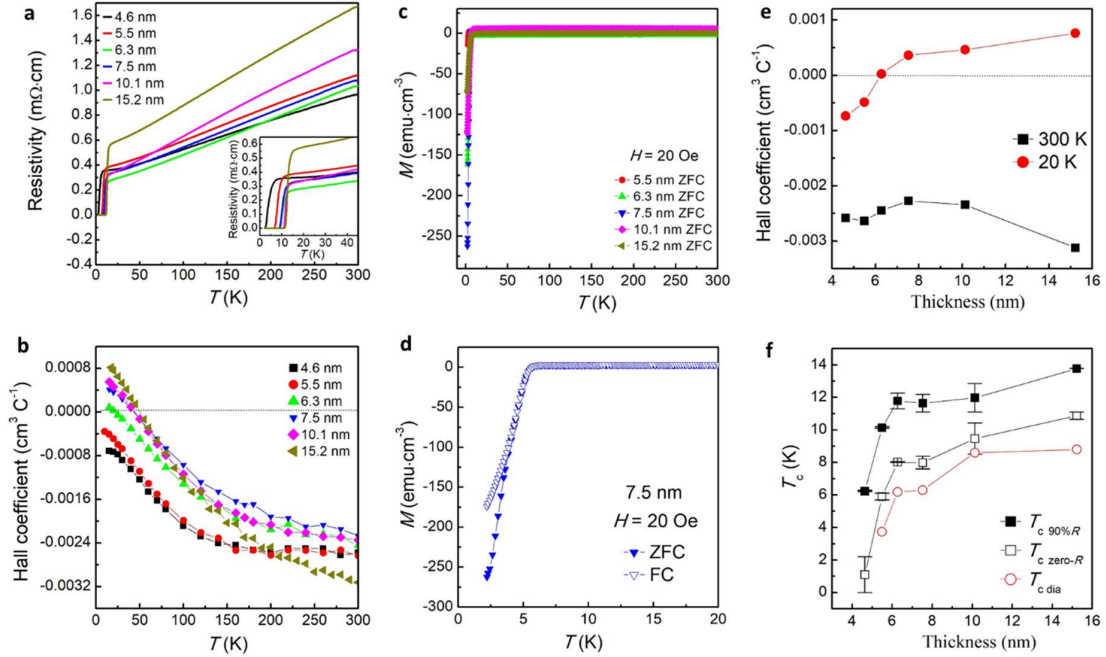

**Figure 2: Thickness dependence of the transition temperature, diamagnetic response, and Hall coefficient.** (a) The resistivity versus temperature ($\rho$-$T$) curves of the $Nd_{0.8}Sr_{0.2}NiO_2$ thin films with different thicknesses from 4.6 to 15.2 nm. The inset shows the zoomed-in $\rho$-$T$ curves at temperatures from 50 to 2 K. (b) The temperature dependence of the normal-state Hall coefficients $R_H$. (c) The temperature dependence of magnetization ($M$-$T$ curve) with zero-field cooling (ZFC) for $Nd_{0.8}Sr_{0.2}NiO_2$ thin films with different thicknesses from 5.5 to 15.2 nm. (d) The zoomed-in $M$-$T$ curves at temperatures from 20 to 2.1 K with field cooling (FC) and ZFC for the sample with a thickness of 7.5 nm. The measurement and cooling fields are 20 Oe. The magnetic field is applied perpendicularly to the $a$-$b$ plane. (e) The $R_H$ at $T = 300$ and 20 K as a function of thickness. (f) The critical temperature, $T_c$, as a function of thickness. The $T_{c,90\%R}$ is defined as the temperature at which the resistivity drops to 90% of the value at 15 K (the onset of the superconductivity). The $T_{c,zero-R}$, is defined as the temperature at which the resistivity drops to be zero and $T_{c,dia}$ is defined as the temperature at the onset of the diamagnetic response. The solid and open square for $T_{c,90\%R}$ and $T_{c,zero-R}$ are the average value of two sets of samples shown in the main text and Supplementary Information, and the error bars represent the variation. The dash lines in (b) and (e) show the position where the $R_H$ is zero.

**Diamagnetic Response**

Figure 2c shows the temperature dependence of magnetization ($M$-$T$ curves) under zero-field cooling (ZFC) mode for $Nd_{0.8}Sr_{0.2}NiO_2$ thin films with thickness varying from 5.5 to 15.2 nm. The normal-state magnetization shows temperature independence. Upon cooling down, the magnetization drops steeply to negative values, confirming the occurrence of superconducting transition in the thin films. Figure 2d shows the zoomed-in *M-T* curves from 20 down to 2.1 K with ZFC and field cooling (FC) for a representative sample with a thickness of 7.5 nm. Similar to the behavior of typical bulk superconductors, a clear diamagnetic transition is seen, and the onset transitions are the same under ZFC and FC modes, unambiguously confirming the existence of the superconducting phase. The ZFC and FC *M-T* curves for other thin films are shown in Supplementary Fig. 2. Supplementary Fig. 3 shows the magnetization measured with the applied magnetic field parallel to the ab plane H || ab (in-plane). Compared to H || c, the change of in-plane magnetization is negligible upon cooling. This is because the film is ultrathin (the maximum thickness of the infinite layer is around 11 nm) and is thinner than the out-of-plane London penetration depth. As the magnetic field is applied parallel to the ab plane, the magnetic field fully penetrates the film, and therefore, the absence of Meissner effect is observed. In certain cases, a small diamagnetic response is observed when H || ab with transition temperature similar to that of measured in H || c, which is likely due to imperfect in-plane alignment of the sample during measurement. The $T_{c,dia}$, defined as the onset transition temperature in diamagnetic response, is shown in Fig. 2f. It is found that $T_{c,dia}$ is slightly lower than $T_{c,zero-R}$, possibly due to filament superconducting domains or presence of inhomogeneity of the superconducting phase in the thin films. Only well below the zero-resistance temperature, the phase coherence occurs in the entire films, and therefore, the Meissner effect is observed. Supplementary Fig. 4a shows the volume susceptibility as a function of temperature, which accounts for the demagnetizing field $H_d = -NM$. A large $\chi_v(T = 2K) < -0.9$ can be observed (for 7.5 nm ZFC MT) which indicates a superconducting volume fraction of more than 90%, suggesting very few non-superconducting unreduced phases or impurities in the sample. In addition, Meissner effect is observed as the negative slope in $M-H$ curve below superconducting temperature, as shown in Supplementary Fig. 4b.

**Electronic Structure**

The unoccupied states of energy bands are crucial to determine the transport properties and could be detected by X-ray absorption spectroscopy (XAS) on oxygen and transition metal edges. To carefully characterize the electronic structures of samples with pure perovskite phase and the resultant infinite-layer phase, we restrict our XAS measurements on thin films with thickness no more than 10.1 nm, as confirmed by XRD measurements. Figure 3a shows the O $K$ edge XAS of 10.1-nm perovskite $Nd_{0.8}Sr_{0.2}NiO_3$ and infinite-layer $Nd_{0.8}Sr_{0.2}NiO_2$ thin films. A prominent pre-peak at ~528.9 eV is observed near the O $K$-edge XAS in $Nd_{0.8}Sr_{0.2}NiO_3$ film, which is attributed to the presence of a ligand hole in oxygen[18-20]. In perovskite nickelates, the oxygen $p$ to nickel $d$ band charge-transfer energy is negative, the electrons spontaneously transfer from oxygen ligands to Ni cations, leaving the holes on the oxygen side even without chemical doping[18]. Excitation of oxygen 1$s$ core electrons to such unoccupied states (holes) give rise to the pre-peak in O $K$-edge XAS. The ligand holes in perovskite $Nd_{0.8}Sr_{0.2}NiO_3$ can also be suggested from the Ni $L_{2,3}$ edge XAS (Fig. 3b), in which a shoulder is visible at ~856 eV corresponding to the electron transition from the Ni core-level 2$p$ to $3d^8\underline{L}$ state ($\underline{L}$ is ligand hole), beside the main sharp peak at ~854.5 eV corresponding to the electron transition from the core-level 2$p$ to $3d^7$ state[18]. As the film is reduced from perovskite to infinite-layer structure, the prominent pre-peak in $Nd_{0.8}Sr_{0.2}NiO_2$ film disappears, which is consistent with the previous study in underdoped $R$NiO$_2$ ($R$ = La, Nd)[19,20]. Instead, another pre-peak with less intensity at a higher energy of ~530.5 eV is observed, suggesting that the oxygen ligand hole is still present in the infinite-layer film. In the Ni $L_{2,3}$ edge XAS (Fig. 3b), the main absorption peaks in both films are observed with a position shift to a lower energy for infinite-layer film, consistent with the reduced Ni valence state as the structures evolve from perovskite to infinite layer[21]. The shoulder is still observed in the infinite-layer $Nd_{0.8}Sr_{0.2}NiO_2$ beside the main peak, even though it shifts to a higher energy position and its intensity is lower compared with that of the perovskite film. This further suggests the presence of oxygen ligand hole state in infinite-layer nickelates.

The parent compound of the canonical cuprate superconductor is a charge-transfer insulator according to the Zaanen-Sawatzky-Allen scheme[22]. The doped holes reside at the oxygen sites due to the strong hybridization of Cu-$3d_{x^2-y^2}$ and O-$2p$ orbitals forming the $3d^9\underline{L}$ states, and therefore, the pre-peak of O $K$ edge emerges upon doping[23]. The spin of the doped holes in oxygen sites and spin in Cu site form the Zhang-Rice singlet state, reducing the cuprate to be an effective single-band system[24]. Whether a similar situation occurs in infinite-layer nickelate, however, is far from clear. The theoretical model suggests that the parent compound $NdNiO_2$ is a Mott-Hubbard insulator, in which the O $2p$ band is below Ni $3d$ lower Hubbard Band[25,26]. This has been suggested by the EELS and XAS measurements that the pre-peak near the O $K$ edge is completely suppressed in undoped $NdNiO_2$ and $LaNiO_2$[19,20]. The pre-peak nature in doped $Nd_{1-x}Sr_xNiO_2$ has not been fully explored by XAS since the STO capping layer obscures the absorption from the underlying nickelate films[27]. Our observation of the obvious pre-peak in O $K$ edge and shoulder in Ni $L_{2,3}$ edge in $Nd_{0.8}Sr_{0.2}NiO_2$ films, which is similar to the cuprate[28], suggests the Ni-O orbital hybridization in infinite-layer nickelate. Note that the EELS measurements have also shown the pre-peak feature in $Nd_{1-x}Sr_xNiO_2$ with increasing doping, although the intensity is weak[20]. Moreover, a prominent pre-peak has also been seen in trilayer nickelates $R_4Ni_3O_8$ (R = La and Pr), which possess the same $NiO_2$ square plane as in the infinite-layer nickelates and an effective 1/3 hole doping[21,29].

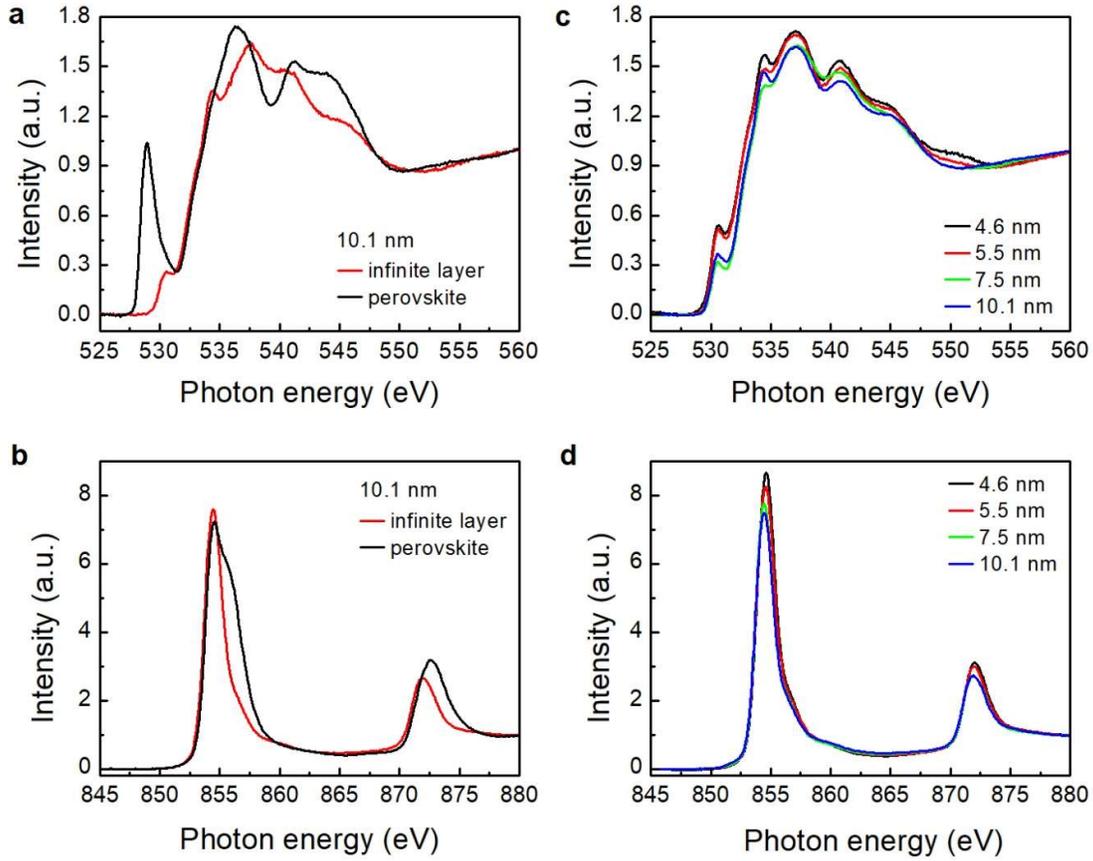

**Figure 3: Thickness dependence of the X-ray Absorption Spectroscopy.** (a) The O $K$ edge and (b) Ni $L_{2,3}$ edge XAS of 10.1 nm perovskite $Nd_{0.8}Sr_{0.2}NiO_3$ and infinite-layer $Nd_{0.8}Sr_{0.2}NiO_2$ thin films. (c) The O $K$ and (d) Ni $L_{2,3}$ edge of infinite-layer $Nd_{0.8}Sr_{0.2}NiO_2$ films with different thicknesses.

Figures 3c and 3d show O $K$ edge and Ni $L_{2,3}$ edge XAS of infinite-layer $Nd_{0.8}Sr_{0.2}NiO_2$ films with different thicknesses. The intensity of pre-peak in O $K$ edge generally increases with decreasing thickness due to the decreasing occupancy (more empty states), suggesting the increase of Ni-O hybridization. The Ni $L_{2,3}$ peak intensity increases and peak position shifts systematically to higher energies as the thickness decreases. This indicates that with decreasing thickness, the Ni 2$p$ core-level electrons have enhanced binding energy, implying a gradual increase of the oxidation state of Ni. The evolution of XAS may be due to the interface and strain effects with decreasing thickness and could be related to the transport properties, as shown above. We note here that the observed XAS signal might contain a minute contribution from the presence of local secondary phases in thick films and/or a parasitic surface layer. Nevertheless, all XAS features are similar to those for the ultrathin single-phase film, suggesting that the contribution from the secondary phase is negligible.

It has been constructed by the bulk electronic structure calculations that the infinite-layer nickelate possesses multiband structures which show the hole pocket originating from a Ni-3$d_{x^2-y^2}$ orbital and the electron pockets from the rare-earth element 5$d_{xy}$ and 5$d_{3z^2-r^2}$ orbitals[2]. Such multiband pictures are consistent with the observed change of $R_H$ in Sr-doped NdNiO$_2$ and PrNiO$_2$[1,3,4,6]. Our observation is that the $R_H$ of Nd$_{0.8}$Sr$_{0.2}$NiO$_2$ also changes with the thickness (Fig. 2b), possibly related to the alteration of band structure due to the interface and strain effects.

**DISCUSSION**

Theoretical calculations suggested the presence of electronic and atomic reconstructions at NdNiO$_2$/SrTiO$_3$ interfaces and the resultant alteration of band structure at the interface[13-16]. At the surface and interface, the NiO$_2$ layer bends and Ni is displaced vertically due to atomic reconstruction and/or the presence of residual apical oxygen at the NdO plane. It is expected that as the thickness decreases, the interface and/or strain effects are more pronounced. Likely, the bending of the NiO$_2$ layer causes the overall increase of the *c*-axis lattice constant, as indicated by the XRD measurement (Fig. 1b). The bending of the NiO$_2$ layer and the resultant tilt of Ni-O bonding causes the extra extraction of the electron from Ni to O. This is consistent with our XAS measurement that the oxidation state of Ni increases as the thickness decreases (Figs. 3b and 3d). Moreover, it has been revealed that the multiband structures become more pronounced at the interface, for example, the mixture of $d_{x^2-y^2}$ and $d_{z^2}$ states are present near Fermi level at the interface, causing extra electron pockets[13-16]. Therefore, as the thickness decreases, the $R_H$ remains negative below 300 K (Fig. 2b). It was proposed that the pairing state of Nd$_{1-x}$Sr$_x$NiO$_2$ changes from (*d*+i*s*)-wave to *d*-wave paring as the doping increases[30], in line with the crossover of $R_H$ from negative to positive sign[3,4]. Interestingly, the interface/surface effect, which caused the negative $R_H$ sign in our result, also induced dominant *s*-wave paring[17] as opposed to *d*-wave symmetry for the bulk[2,31]. Overall, even though the Meissner effect is confirmed in the nickelate thin films, the interfacial effects due to the atomic reconstruction play an important role in the modification of multiband structures.

**METHODS**

**Thin Film Growth and Reduction**

The perovskite Nd$_{0.8}$Sr$_{0.2}$NiO$_3$ thin films with different thicknesses were grown on a TiO$_2$-terminated (001) SrTiO$_3$ (STO) substrate using a pulsed laser deposition (PLD) technique with a 248-nm KrF excimer laser. No capping layer is introduced for all samples. The deposition temperature and oxygen partial pressure $P_{O_2}$ for all samples were 600 °C and 150 mTorr, respectively. The laser energy density on the target surface was set to be 1.8 Jcm$^{-2}$. After deposition, the samples were annealed for 10 min at 600 °C and 150 mTorr and then cooled down to room temperature at a rate of 8 °C/min. The as-grown samples were cut into pieces with a size of around 2.5 × 5 mm². The pieces were then embedded with about 0.15 g of $CaH_2$ powder and wrapped in aluminum foil, and then placed into the PLD chamber for reduction. Using the PLD heater, the wrapped samples were heated to $340 - 360$ °C at a rate of 25 °C/min and kept for 80 minutes, and then cooled down to room temperature at a rate of 25 °C/min.

**Electrical and Magnetic Characterizations**

The wire connection for the electrical transport measurement was made by Al ultrasonic wire bonding. The electrical transport and magnetization measurements were performed using a Quantum Design Physical Property Measurement System and Superconducting Quantum Interference Device Magnetometer, respectively.

**X-ray Diffraction and Absorption Spectrocopy**

The X-ray diffraction (XRD) measurement was done in the X-ray Diffraction and Development (XDD) beamline at Singapore Synchrotron Light Source (SSLS) with an X-ray wavelength of $\lambda = 1.5404$ Å. The XAS measurements were performed using linearly polarized X-ray from the Soft X-ray-ultraviolet (SUV) beamline at SSLS, using a total electron yield (TEY) detection method. The incidence angle $(90 - \theta)°$ of X-rays refers to the normal of the sample surface, which was varied by rotating the polar angle of the sample. The spectra were measured in a grazing-incident alignment ($\theta = 20°$) to obtain better sample signals. The spectra were normalized to the integrated intensity at the tail of the spectra after subtracting an energy-independent background.

**Scanning Transmission Electron Microscopy**

The high-angle annular dark-field scanning transmission electron microscopy (HAADF-STEM) imaging was carried out at 200 kV using a JEOL ARM200F microscope, and the cross-sectional TEM specimens were prepared by a focused ion beam machine (FEI Versa 3D). All the data of the same thickness are measured from the exact same sample to ensure consistency in comparing the results of various measurements.

**DATA AVAILABILITY**

Data available on request from the authors

**ACKNOWLEDGMENTS**

This research is supported by the Agency for Science, Technology, and Research (A*STAR) under its Advanced Manufacturing and Engineering (AME) Individual Research Grant (IRG) (A1983c0034) and the Singapore National Research Foundation (NRF) under the Competitive Research Programs (CRP Grant No. NRF-CRP15-2015-01). P. Y. is supported by Singapore Synchrotron Light Source (SSLS) via NUS Core Support C-380-003-003-001. The authors would also like to acknowledge the SSLS for providing the facility necessary for conducting the research. The Laboratory is a National Research Infrastructure under the National Research Foundation (NRF) Singapore.## AUTHOR CONTRIBUTIONS

SWZ and AA conceived the project. SWZ prepared the thin films and conducted the electrical and magnetic measurements with the assistance from KH, ZH, YC, LEC, DYW, ZTZ and ZSL. SWZ, PY, CST and XMY conducted the XRD measurements. XMY, CZD, CST and ATSW conducted the XAS measurements. CJL and SJP conducted the STEM measurements. SWZ and AA wrote the manuscript with contributions from all authors. All authors have discussed the results and the interpretations.

## COMPETING INTERESTS

The authors declare no competing interests

# Supplementary Information

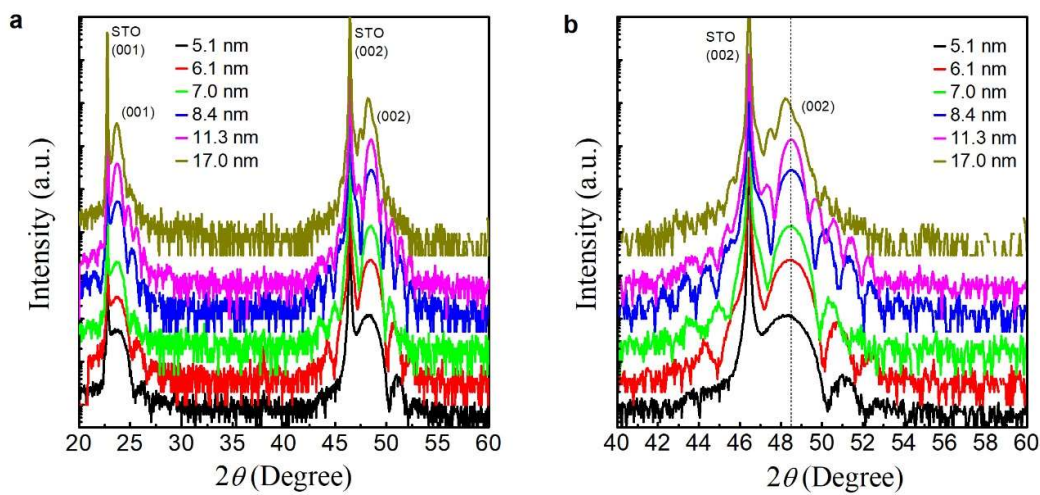

**Supplementary Figure 1:** (a) The XRD $\theta$-$2\theta$ scan patterns of the as-grown $Nd_{0.8}Sr_{0.2}NiO_3$ thin films with different thicknesses on STO substrates. The intensity is vertically displaced for clarity. Only the (00$l$) perovskite peaks are observed, where $l$ is an integer, confirming the $c$-axis oriented epitaxial growth. (b) The zoomed-in XRD $\theta$–$2\theta$ scan patterns at angles from 40 to 60 degrees. The black dash line shows the position of the (002) perovskite peaks for thin films with different thicknesses. One can see that for thin films with thicknesses from 5.1 to 11.3 nm, the (002) peaks are at the same position, indicating the same $c$-lattice constant. For the film with a thickness of 17 nm, the presence of multiple peaks is seen at (002) position, suggesting the presence of mixed phases as the film thickness is increased, consistent with previous reports. The film thickness is calculated by fitting the Laue fringes in the vicinity of the (002) peak.

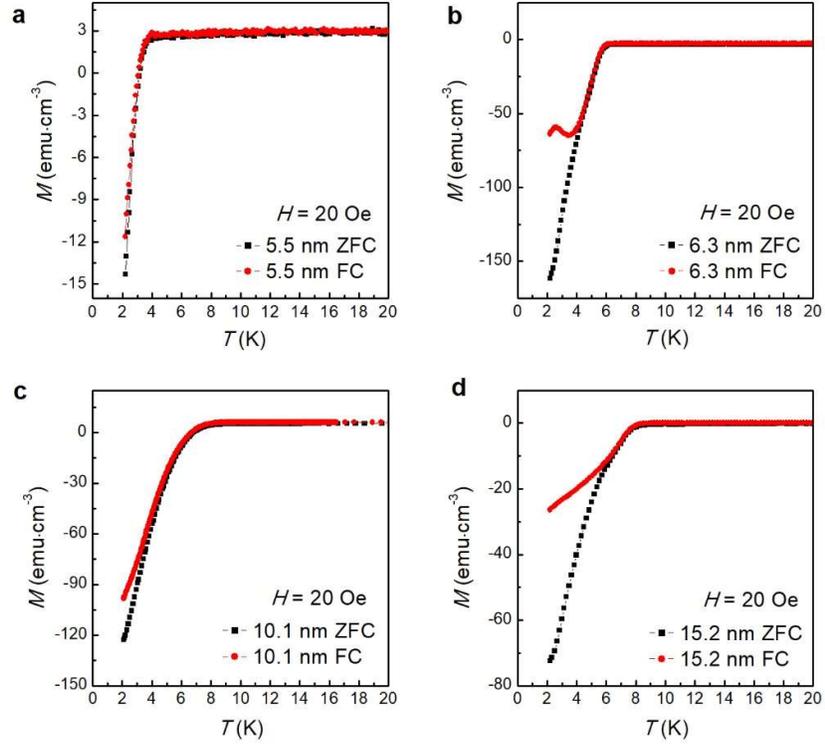

**Supplementary Figure 2:** Temperature dependence of magnetization (*M-T* curve) with zero-field cooling (ZFC) and field cooling (FC) for Nd$_{0.8}$Sr$_{0.2}$NiO$_2$ thin films with different thickness of (a) 5.5 nm, (b) 6.3 nm, (c) 10.1 nm and (d) 15.2 nm. The *M-T* curves are shown at temperatures from 20 to 2.1 K. The measurement and cooling fields are 20 Oe. The magnetic field is applied perpendicularly to the *a-b* plane.

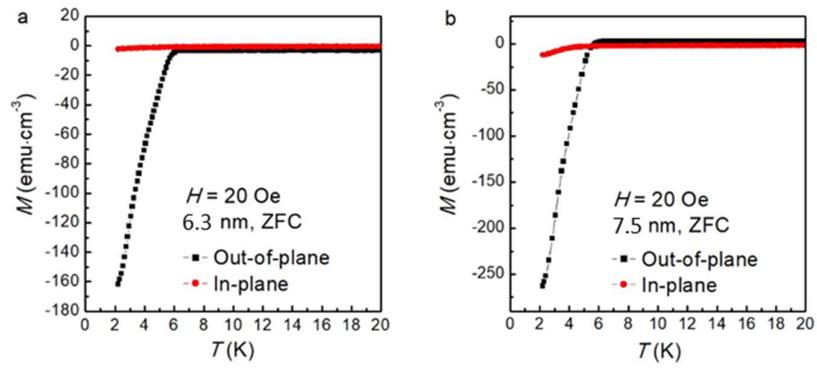

**Supplementary Figure 3:** Temperature dependence of magnetization with zero-field cooling (ZFC) mode for $Nd_{0.8}Sr_{0.2}NiO_2$ films with thickness of 6.3 nm and 7.5 nm. The magnetizations are shown for magnetic field perpendicular (out-of-plane) and parallel (in-plane) to the ab plane.

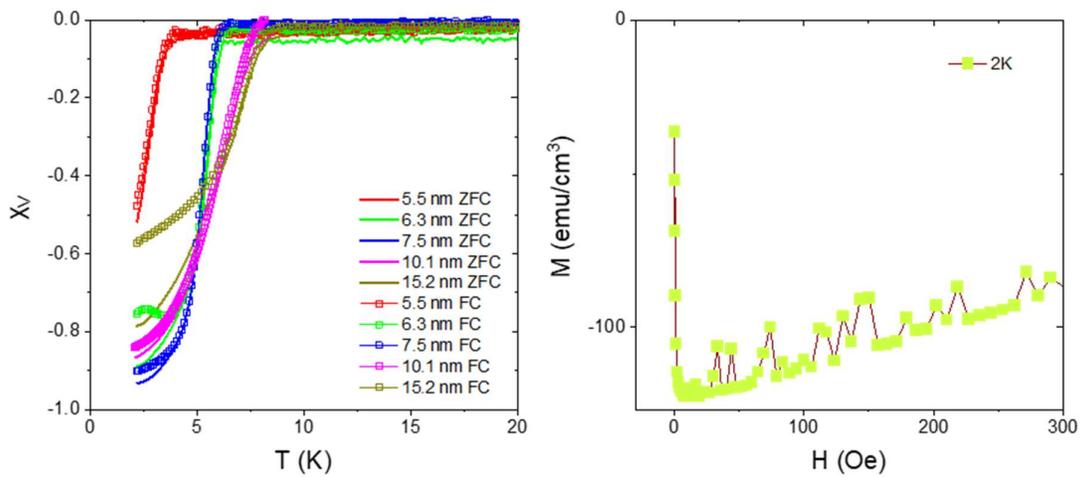

**Supplementary Figure 4: (Left)** Volume susceptibility as a function of temperature calculated from susceptibility measurement under H || c = 20 Oe in ZFC and FC conditions. In S.I. unit, $\chi_V = -1$ indicates 100% superconducting volume fraction (perfect diamagnetism). **(Right)** Magnetisation as a function of applied magnetic field at the out-of-plane direction in ZFC conditions. The STO substrate diamagnetic contribution is subtracted from the raw data with linear fitting of the gradient at large magnetic field range.

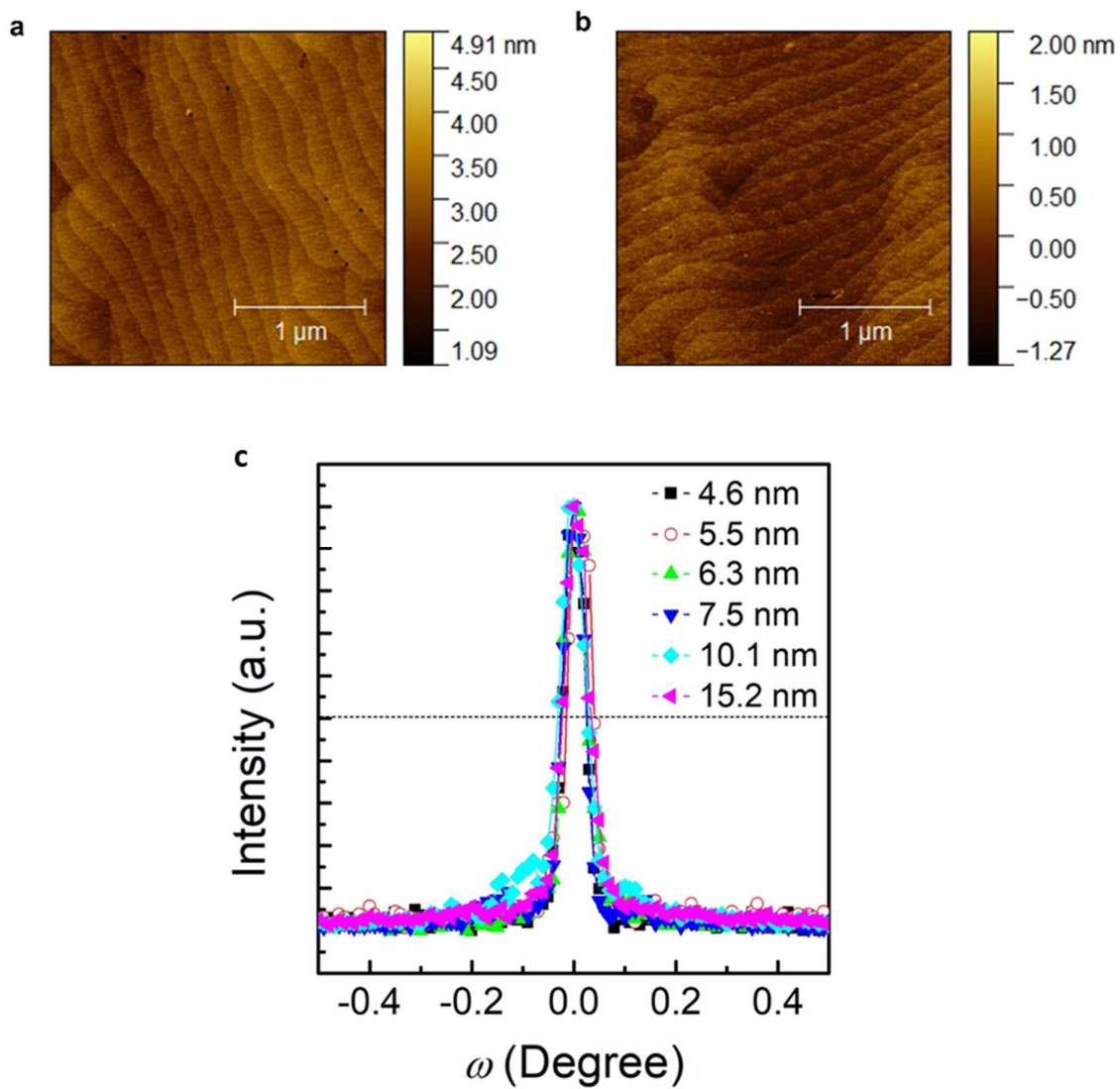

**Supplementary Figure 5: (a-b)** AFM images of the 6.3 nm sample surface (a) before (b) after reduction. **(c)** The full width at half-maximum (FWHM) of the (002) rocking curves. The value of FWHM is less than 0.06, indicating a good quality of the infinite-layer film.

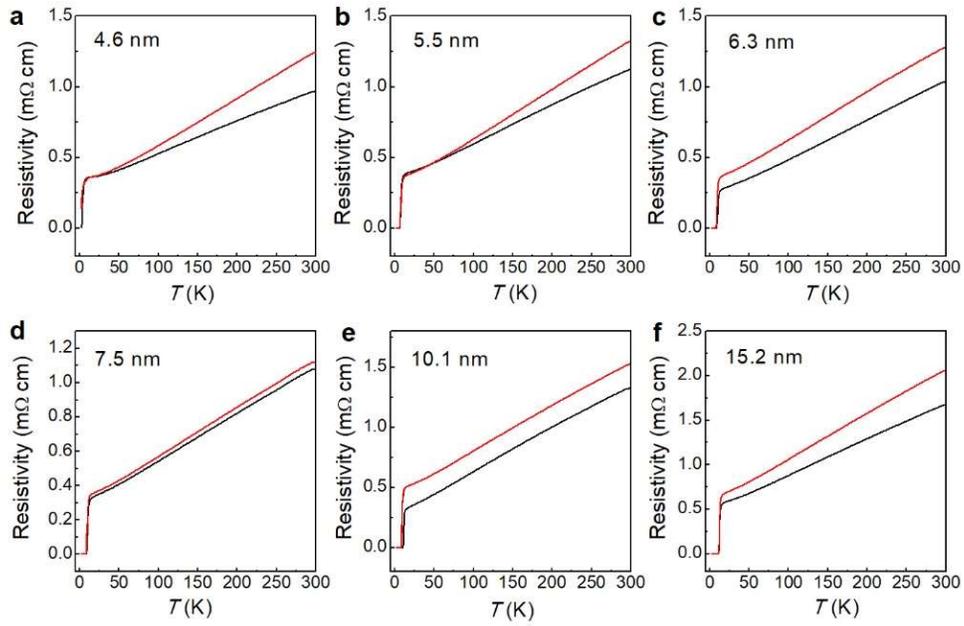

**Supplementary Figure 6:** The resistivity versus temperature ($\rho$-$T$) curves another set of samples. The black curves are the data shown in the main text, and the red curves are for another set of samples.